# High Spatial and Temporal Resolution NIR-IIb Gastrointestinal Imaging in Mice

Chao Mi[1], Ming Guan[2,*], Xun Zhang[2], Liu Yang[2], Sitong Wu[2], Zhichao Yang[2], Zhiyong Guo[2], Jiayan Liao[3], Jiajia Zhou[3], Dayong Jin[2,3,*], Xiaocong Yuan[1,*]

[1]Nanophotonics Research Center, Shenzhen Key Laboratory of Micro-Scale Optical Information Technology & Institute of Microscale Optoelectronics, Shenzhen University, Shenzhen, China

[2]UTS-SUStech Joint Research Centre for Biomedical Materials and Devices, Department of Biomedical Engineering, Southern University of Science and Technology, Shenzhen, Guangdong 518055, P. R. China

[3]Institute for Biomedical Materials and Devices (IBMD), Faculty of Science, University of Technology Sydney, NSW 2007, Australia

*Correspondence: guanm@mail.sustech.edu.cn; jindy@sustech.edu.cn; xcyuan@szu.edu.cn

## Abstract

Conventional biomedical imaging modalities, including endoscopy, X-rays, and magnetic resonance, are invasive and cannot provide sufficient spatial and temporal resolutions for regular imaging of gastrointestinal (GI) tract to guide prognosis and therapy of GI diseases. Here we report a non-invasive method for optical imaging of GI tract. It is based on a new type of lanthanide-doped nanocrystal with near-infrared (NIR) excitation at 980 nm and second NIR window (NIR-IIb) (1500~1700 nm) fluorescence emission at around 1530 nm. The rational design and controlled synthesis of nanocrystals with high brightness have led to an absolute quantum yield (QY) up to 48.6%. Further benefitting from the minimized scattering through the NIR-IIb window, we enhanced the spatial resolution by 3 times compared with the other NIR-IIa (1000~1500 nm) contract agents for GI tract imaging. The approach also led to a high temporal resolution of 8 frames per second, so that the moment of mice intestinal peristalsis happened in one minute can be captured. Furthermore, with a light-sheet imaging system, we demonstrated a three-dimensional (3D) imaging of the stereoscopic structure of the GI tract. Moreover, we successfully translate these advances to diagnose inflammatory bowel disease (IBD) in a pre-clinical model of mice colitis.

# Introduction

Digestive disorders encompass a great variety of diseases ranging from mild to severe[1,2]. For example, as a common chronic and recurrent GI disease, inflammatory bowel disease (IBD) is closely related to colorectal cancer, which accounts for one one-sixth of ulcerative colitis-related deaths[3,4]. Several imaging modalities are used for assessing GI conditions in the clinic, such as endoscopy imaging[2,5], X-ray imaging[6], computed tomography imaging[7], and magnetic resonance imaging[8]. Nonetheless, current techniques are invasive and have limited spatial and temporal resolutions to visualize the details of GI tract and its motor patterns. Hence, GI peristalsis and segmentation usually require *ex vivo* measurements[9].

Fluorescence imaging provides an alternative tool for non-invasive imaging modalities. Furthermore, due to the advantages of low photon scattering and autofluorescence for deep tissue penetration and high spatial resolution, fluorescence imaging in the second-near infrared window (NIR-II, 1000~1700 nm) has recently attracted extensive attention[10–16]. Thus, NIR-II imaging has been applied for *in vivo* visualizations of the vasculature, organs, tumors, and GI tract[17–21]. In 2017, Zhang's group reported the NIR-II (~1060 nm) imaging of drug release in the GI tract by oral administration of $Nd^{3+}$ doped nanoparticles[22]. In 2019, Hong et al. studied the NIR-II (~1034 nm) real-time imaging of normal gastrointestinal motility and intestinal obstruction in mouse by organic aggregation-induced emission fluorophore[23]. In 2020, Li et al. demonstrated the NIR-II (~1050 nm) imaging of the GI tract by engineering a protein corona structure consisting of ribonuclease-A and Au nanoclusters[24]. Recently, the spectral computed tomography was combined with the NIR-II (~1065 nm) imaging modalities to visualize the *in vivo* GI tract[7].

According to Mie theory[25,26], the optical window at NIR-IIb has minimal photon scattering when through the deep tissue, which can optimize the imaging resolution and signal-to-noise ratio, compared with the NIR-IIa window. However, the tissue absorption at NIR-IIb bandwidth becomes relatively strong[25,27], which puts forward higher requirements on the high brightness of emitted fluorescence. In 2019, an ultra-bright $Er^{3+}$-doped $NaYbF_4$: $Er^{3+}$, $Ce^{3+}$, $Zn^{2+}$@$NaYF_4$ nanocrystal with ~1530 nm emission was reported for NIR-IIb imaging of tumors[28]. In a parallel investigation, a type of $Er^{3+}$-doped $NaCeF_4$: $Yb^{3+}$, $Er^{3+}$ nanocrystals with a high QY of 32.8% for NIR-IIb emission was studied[29]. In addition, the $Er^{3+}$-doped nanocrystals exhibit a high degree of photochemical stability and highly controllable morphology[30,31,32], which holds a great promise for high-resolution NIR-IIb GI tract imaging.

Here, we report a non-invasive NIR-IIb approach with sufficient spatial and temporal resolutions for GI tract imaging and GI disease diagnosis. We first designed and synthesized the $NaYbF_4$: $Er^{3+}$, $Ce^{3+}$, $Zn^{2+}$@$CaF_2$ (ErNCs) core-shell nanocrystals with strong NIR-IIb fluorescence at ~1530 nm. By coating the nanocrystals with gum arabic (GA) to obtain ErNCs@GA, we demonstrated a NIR-IIb contrast meal for real-time high resolution planar and 3D intestine imaging. Furthermore, by coating carboxymethylcellulose sodium (CMC-Na) onto ErNCs, we achieved another contrast meal, named ErNCs@CMC-Na complex, and successfully diagnosed IBD in mice.

**Results and Discussion**

**Design and synthesis of NIR-IIb emissive nanocrystals with remarkable high quantum yield, stability, and biocompatibility.** As passivating nanocrystals with an inert shell can largely enhance the luminescent efficiency[33,34], we designed and synthesized a series of cubic phase $NaYbF_4$: 2% $Er^{3+}$, 2% $Ce^{3+}$, 10% $Zn^{2+}$ nude core with different sizes and corresponding $CaF_2$ passivated core-shell nanocrystals through a multi-step epitaxial seeded growth method[28,35], and the NIR-IIb emission from $Er^{3+}$ is enhanced by blocking the surface quenching (Figure 1a). Knowing that calcium and fluoride ions are common endogenous components and lattice substituents of calcified tissues, such as bones and teeth, the passive shell of $CaF_2$, with a low lattice mismatch with the cubic phase $NaYbF_4$ host, makes the core-shell nanoparticles high crystallizability and biocompatibility[36]. The TEM images in Figure 1b show the nude nanoparticles have a uniform morphology with diameters of 11.1 nm, 15.2 nm, 43.8 nm, and 65.6 nm, respectively, while the sizes of corresponding core-shell nanoparticles are 19.1 nm, 27.8 nm, 52.5 nm, and 80.9 nm, respectively (Figure S1). The thicknesses of the $CaF_2$ passivated shells are determined around 4 nm, 6.3 nm, 4.4 nm, and 7.6 nm, respectively. More characterization tests including the XRD, EDS, element mapping, and the ~1550 nm florescence lifetime measurements are given in Figure S2-S5. These results can verify the cubic-phase crystal lattice, the containing chemical elements, and the core-shell structure of the prepared ErNCs.

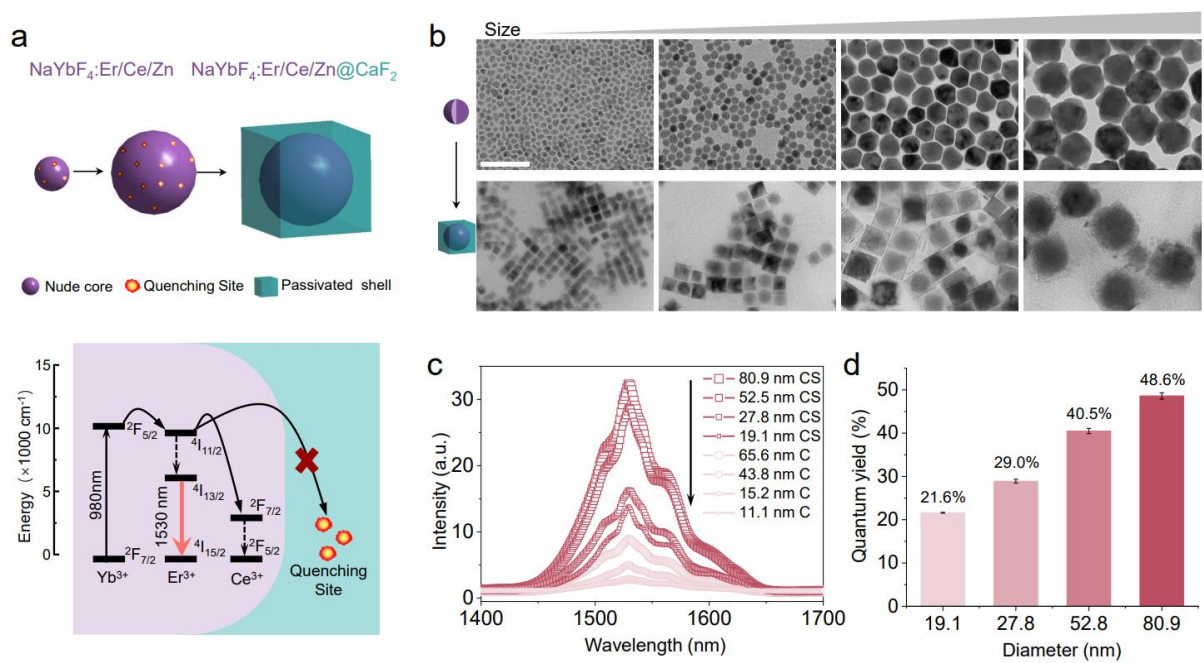

**Figure 1. NIR-IIb fluorescent nanocrystals with high quantum yield.** (a) The structure design of ultra-bright NIR-IIb emissive ErNCs, and the energy level diagram depicts the passivation effect and enhanced

NIR-IIb emission. (b) TEM images of nude core nanoparticles (upper panel) showing the size increasing from 11.1 nm to 15.2 nm, 43.8 nm and 65.6 nm, and the size of the corresponding core-shell nanoparticles (lower panel) increasing from 19.1 nm to 27.8 nm, 52.5 nm and 80.9 nm, respectively. (c) NIR-IIb emissive spectra of core and core-shell nanocrystals in (b) (C: Core nanocrystals; CS: Core-shell nanocrystals). (d) The QYs at ~1530 nm emissions of the corresponding core-shell NIR-IIb ErNCs in (b).

The spectra in Figure 1c confirm the enhanced ~1530 nm NIR-IIb emissions at a larger size and with $CaF_2$ shell. Upon a 980 nm excitation with the power density of 100 mW/cm$^2$, the NIR-IIb emission's QY at ~1530 nm for the core-shell nanoparticles with sizes of 19.1 nm, 27.8 nm, 52.5 nm and 80.9 nm in diameter was determined to be 21.6%, 29.0%, 40.5%, and 48.6%, respectively (Figure 1d). To the best of our knowledge, QY of 48.9% is the highest value among $Er^{3+}$-activated NIR-IIb nanoparticles[11].

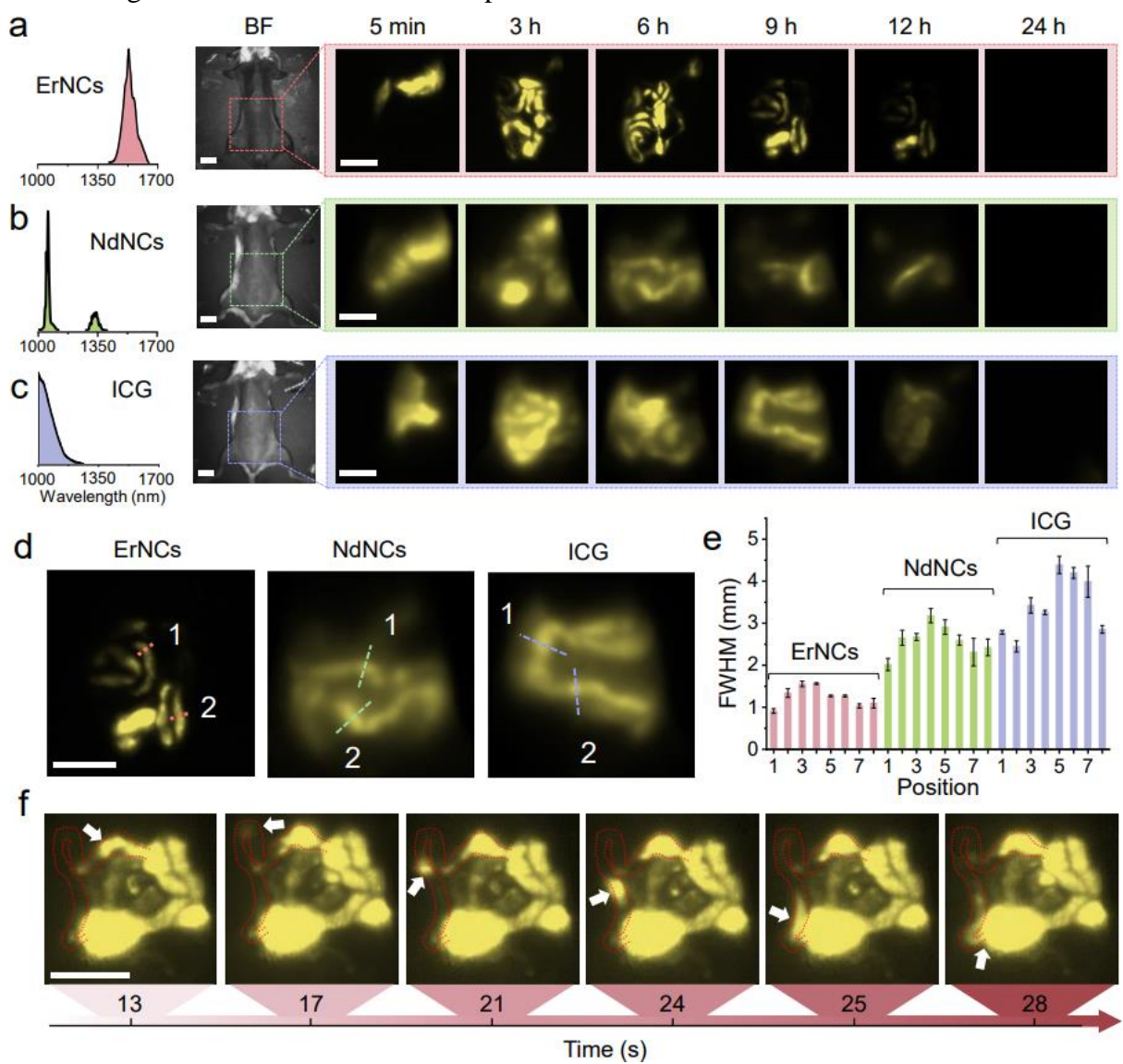

**Figure 2. *In vivo* NIR-IIb imaging on GI tract and intestinal peristalsis.** (a)(b)(c) Time-course imaging, performed at 5 minutes, 3, 6, 9, 12 and 24 hours post the gavage, to show the enhanced spatial resolution achieved by 1530 nm NIR-IIb contrast meal ErNCs@GA (a), compared with ~1000 nm NIR-IIa emissive

NdNCs (b) and ICG (c). The spectra represent the NIR-II emissions of ErNCs, NdNCs and ICG, respectively. (min: minutes, h: hours, BF: bright field). (d)(e) Quantitative analysis on the FWHM (e) of 8 different imaging positions indicated by the dash lines in Figure 2 (d) and S12. The locations were chosen from 4 mice models for each contrast meal. Red: ErNCs; Green: NdNCs; Blue: ICG. (f) The high temporal resolution of NIR-IIb imaging for real-time monitoring of the dynamics of GI peristalsis and meal flow in the intestinal tract. The white arrows in (f) show the peristalsis and meal flow direction.

**Long-term NIR-IIb *in vivo* imaging of GI tract in mice**. We developed a NIR-IIb ErNCs@GA contrast meal by mixing ErNCs with GA, as GA is a plant-derived viscosity-enhancing polysaccharide and has been widely used as the emulsifier[6,37]. In Figure 2a, after oral administration of 100 μL ErNCs@GA contrast meal, we conducted the time-course intestine imaging of mice by a purpose-built *in vivo* NIR-II imaging system (Figure S6). Owing to the strong NIR-IIb emission around 1530 nm, intense signals first appeared in the mice's stomach in the initial 5 minutes after oral administration. Then the contrast meal gradually moved to the intestine and reached the maximum value after 3 hours. Because of the low tissue scattering of ~1530 nm emissions, NIR-IIb imaging precisely resolved the location and outline of the intestine, rectum, and colon through the intact mice skin and soft tissue. Moreover, the well adhesion ability of GA to the intestinal mucosa extended the *in vivo* imaging window for over 10 hours. After 24 hours, the NIR-IIb signals completely disappeared, as the contrast meal have been excreted by feces (Figure S7), and the hematoxylin and eosin (H&E) staining of the sectioned tissues and *ex vivo* NIR-IIb signal examination on the mice organs further approved that the NIR-IIb contrast meal is harmless (Figure S8 and S9).

To demonstrate the enhanced spatial resolution at the NIR-IIb window, we compared the imaging results at the NIR-IIa window by using $NaYF_4$: $Nd^{3+}$ nanocrystals (NdNCs) in Figure 2b and S10 with strong emissions around 1064 nm, and the clinically approved ICG dye with emissions between 1000-1200 nm in Figure 2c [38,39]. The results showed neither led to sufficient spatial resolution images of the GI tract. The quantitative analysis of full width at half maximum (FWHM) further confirmed that the *in vivo* imaging resolution by ErNCs@GA was ~3 times higher than that by NIR-IIa contrast meals (Figure 2d, 2e, S11, S12). More results in Figures S13 to S15 show the stronger biological tissue scattering at a shorter fluorescent wavelength range.

The high brightness of NIR-IIb contrast meal ErNCs@GA and the minimized scattering effect through tissues further allow us to monitor GI motions and transit of the meal in the intestinal tract with a video rate of 8 frames per second (Figure 2f and SI video). Significantly, we can clearly capture the intermittent flow of the contrast meal from the small intestine to the cecum within one minute as well as GI peristalsis (SI video), with distinct dynamic details for understanding the patterns of motions.

**3D imaging of GI tract by NIR-IIb light-sheet illumination.** The NIR-IIb contrast meal with high brightness further allows a purpose-built NIR-IIb system with an orthogonally arranged light-sheet illumination (Figure 3a-c) to build the 3D structure of the GI tract (Figure 3d)[40]. Our system can perform high-speed optical sectioning of up to ~6 mm deep into the tissue, as shown in the y and z-section image in Figure 3e. The advance provides the 3D spatial information of the GI tract by combining our NIR-IIb light-sheet imaging system with the bright NIR-IIb contrast meal.

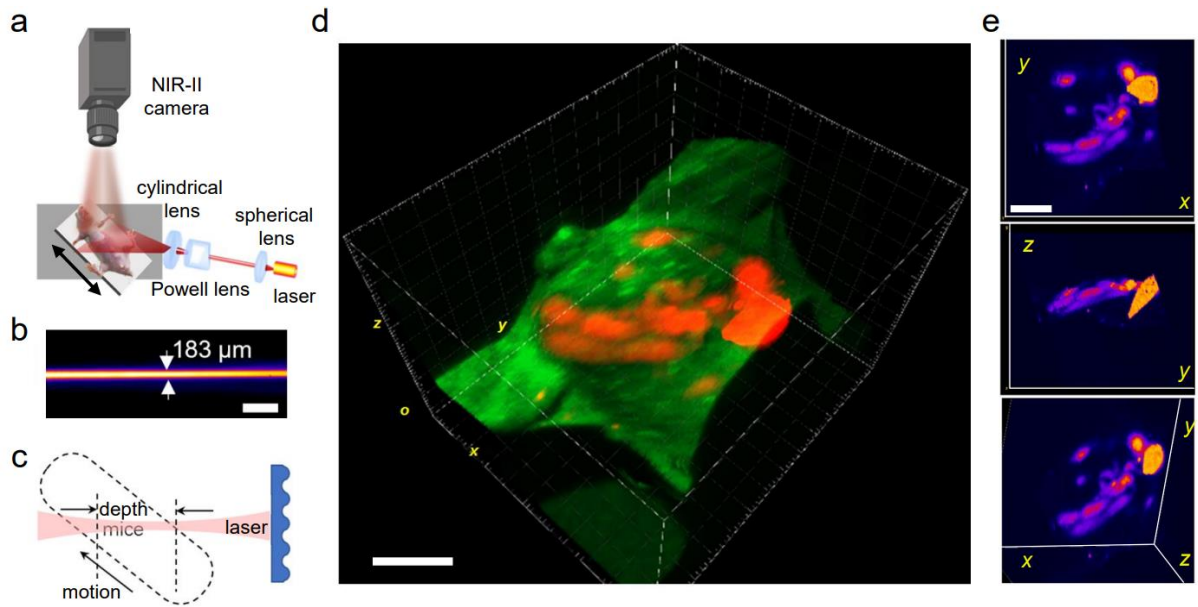

**Figure 3. Three-dimensional imaging of GI tract by a light-sheet illumination system.** (a) Schematic illustration of a NIR-IIb light-sheet imaging system. (b) The cross-section of the light-sheet illumination beam. Scale bar: 1 mm. (c) Schematic of optical sectioning through deep tissue. (d) The 3D NIR-II imaging of GI tract by using the contrast meal ErNCs@GA. Green: Skin; Red: GI tract. Scale bar: 10 mm. (e) Optical sections of the 3D image from different angles of view. Scale bar: 10 mm.

**Diagnosis of inflammatory bowel disease (IBD) by *in vivo* NIR-IIb imaging in mice.** Next, we prepared a contrast meal ErNCs@CMC-Na (Figure 4a) to demonstrate our NIR-IIb imaging system for IBD diagnosis[41]. A pre-clinical model of ulcerative colitis (UC), a type of IBD, was built by oral administration of dextran sodium sulfate (DSS) solution to mice for 7 days [42,43] (Figure 4b). The H&E colon staining result, after NIR-IIb imaging experiments, confirms the successful establishment of ulcerative colitis after DSS administration, showing severe erosion, crypt destruction, and infiltration of leukocytes (Figure S16). On day 8, after the oral administration of ErNCs@CMC-Na, the control group of healthy mice showed clear NIR-IIb images on the GI tract for the first 6 hours, while the signals from the 8 day (d)-UC group lasted for over 24 hours, as a clear contrast, due to the presence of UC. The long duration of NIR-IIb signals for the UC model mice is due to the robust adhesion of ErNCs@CMC-Na complex to the inflamed intestine tissue. It is worthy noting that the ErNCs@CMC-Na can be excreted more quickly than ErNCs@GA, as the signals from the normal GI tract last for a shorter duration of ~6 hours, thus the ErNCs@CMC-Na is more suitable for the rapid IBD diagnosis than ErNCs@GA.

Moreover, we have treated the UC model mice by the oral administration of 5-aminosalicylic acid (5-ASA, an anti-inflammatory agent) from day 8 to day 14. On day 15, assisted by the intake of ErNCs@CMC-Na contrast meal, similarly, the signals of GI tract images from the treated group only lasted for 9 hours, as a sign of successful treatment (Figure 4c). As a comparison, the 15 d-UC group without medication displayed a series of obvious NIR-IIb signals after 24 hours from the digestive tract. More cases on the UC diagnosis can be found in Figures S17 to S20. The *ex vivo* NIR-IIb imaging on the harvested mice colons from the four groups (Figure S21) confirms the diagnosed results from *in vivo* NIR-IIb imaging experiments above.

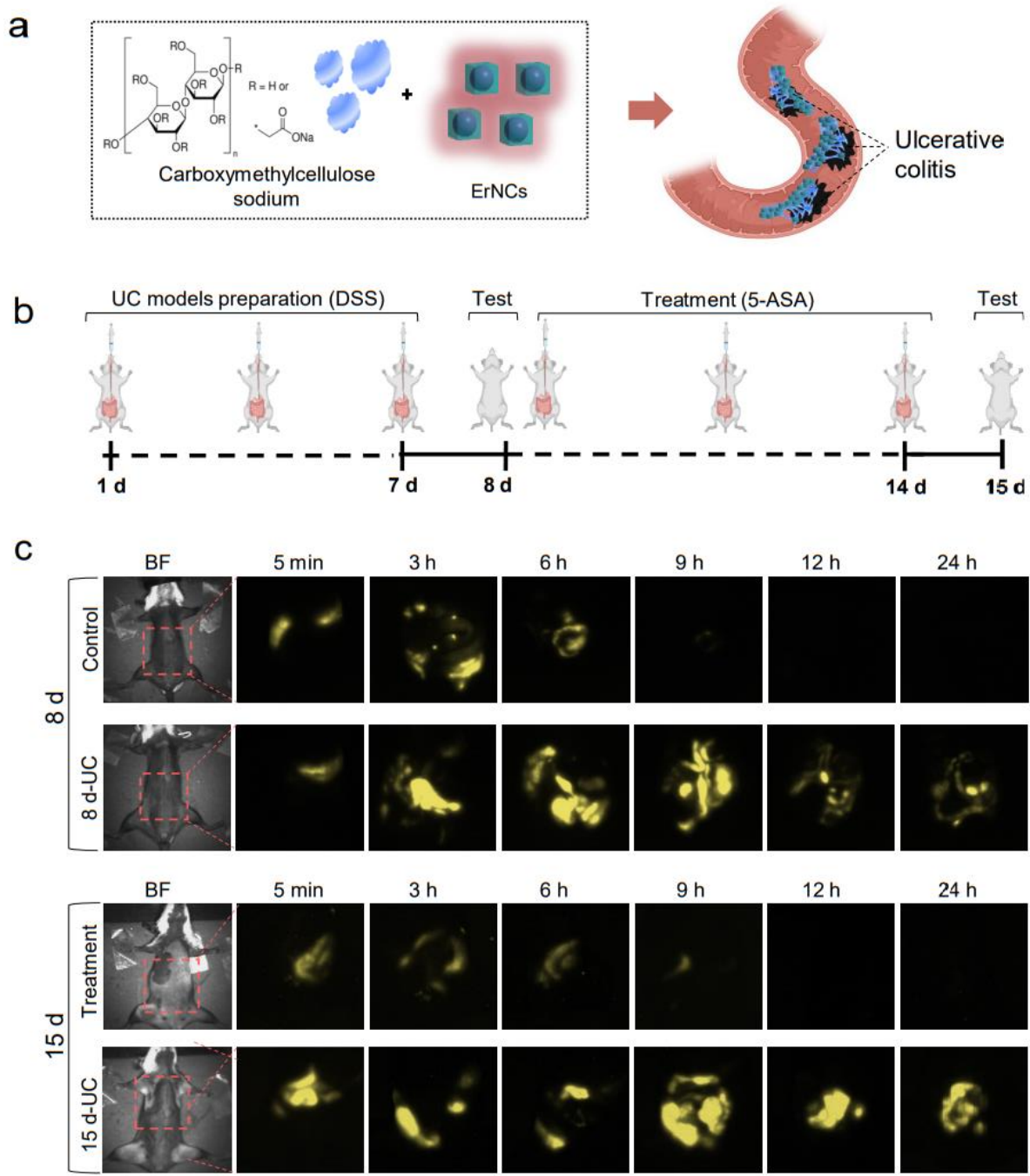

**Figure 4. The NIR-IIb imaging to diagnose and real-time monitor the treatment of ulcerative colitis (UC).** (a) Schematic on the design of NIR-IIb contrast meal. (b) Timeline of UC model establishment and treatment. (c) NIR-IIb images of the control group (healthy mice), the 8 day (d)-UC group, the treatment group, and the 15 d-UC group, collected at the time series of 5 min, 3, 6, 9, 12, and 24h after gavage of NIR-IIb contrast meal.

## Conclusion

We developed a type of ErNCs that emits strong emissions at the NIR-IIb window with a quantum yield of up to 48.9%. We modified ErNCs as an intestinal contrast meal for the real-time imaging of stereoscopic GI tract, intestinal peristalsis, as well as diagnosis and monitoring the treatment of IBD. In comparison to NIR-IIa contrast meals, by using NIR-IIb contrast meals, this work achieved high spatial and temporal resolutions in imaging intestine, rectum, and colon in mice, and with excellent photostability under the continuous laser irradiation. With these advances, we successfully diagnosed IBD and evaluated the treatment efficacy towards pre-clinical translation.

**Acknowledgments**
The authors thank Fulin Lin (Xiamen Institute of Rare Earth Materials, Haixi Institutes, Chinese Academy of Sciences) for the quantum yield measurements. The authors acknowledge the financial support from the National Natural Science Foundation of China (C. Mi, 62005179), China Postdoctoral Science Foundation (C. Mi, 2020M682866), Major International (Regional) Joint Research Project of NSFC (51720105015), the Shenzhen Science and Technology Program (KQTD20170810110913065; 20200925174735005).

# References

(1) Kim, E. R.; Chang, D. K. Colorectal Cancer in Inflammatory Bowel Disease: The Risk, Pathogenesis, Prevention and Diagnosis. *World J. Gastroenterol.* **2014**, *20* (29), 9872–9881. https://doi.org/10.3748/wjg.v20.i29.9872.

(2) Yim, J. J.; Harmsen, S.; Flisikowski, K.; Flisikowska, T.; Namkoong, H.; Garland, M.; Van Den Berg, N. S.; Vilches-Moure, J. G.; Schnieke, A.; Saur, D.; Glasl, S.; Gorpas, D.; Habtezion, A.; Ntziachristos, V.; Contag, C. H.; Gambhir, S. S.; Bogyo, M.; Rogalla, S. A Protease-Activated, near-Infrared Fluorescent Probe for Early Endoscopic Detection of Premalignant Gastrointestinal Lesions. *Proc. Natl. Acad. Sci. U. S. A.* **2021**, *118* (1), 1–8. https://doi.org/10.1073/pnas.2008072118.

(3) Jess, T.; Loftus, E. V.; Velayos, F. S.; Harmsen, W. S.; Zinsmeister, A. R.; Smyrk, T. C.; Schleck, C. D.; Tremaine, W. J.; Melton, L. J.; Munkholm, P.; Sandborn, W. J. Risk of Intestinal Cancer in Inflammatory Bowel Disease: A Population-Based Study From Olmsted County, Minnesota. *Gastroenterology* **2006**, *130* (4), 1039–1046. https://doi.org/10.1053/j.gastro.2005.12.037.

(4) Axelrad, J. E.; Lichtiger, S.; Yajnik, V. Inflammatory Bowel Disease and Cancer: The Role of Inflammation, Immunosuppression, and Cancer Treatment. *World J. Gastroenterol.* **2016**, *22* (20), 4794–4801. https://doi.org/10.3748/wjg.v22.i20.4794.

(5) Molckovsky, A.; Wong Kee Song, L. M.; Shim, M. G.; Marcon, N. E.; Wilson, B. C. Diagnostic Potential of Near-Infrared Raman Spectroscopy in the Colon: Differentiating Adenomatous from Hyperplastic Polyps. *Gastrointest. Endosc.* **2003**, *57* (3), 396–402. https://doi.org/10.1067/mge.2003.105.

(6) Shahidi, S.; Iranpour, S.; Iranpour, P.; Alavi, A. A.; Mahyari, F. A.; Tohidi, M.; Safavi, A. A New X-Ray Contrast Agent Based on Highly Stable Gum Arabic-Gold Nanoparticles Synthesised in Deep Eutectic Solvent. *J. Exp. Nanosci.* **2015**, *10* (12), 911–924. https://doi.org/10.1080/17458080.2014.933493.

(7) Zhuang, P.; Xiang, K.; Meng, X.; Wang, G.; Li, Z.; Lu, Y.; Kan, D.; Zhang, X.; Sun, S. K. Gram-Scale Synthesis of a Neodymium Chelate as a Spectral CT and Second near-Infrared Window Imaging Agent for Visualizing the Gastrointestinal Tractin Vivo. *J. Mater. Chem. B* **2021**, *9* (9), 2285–2294. https://doi.org/10.1039/d0tb02276d.

(8) Ng, C.; Dellschaft, N. S.; Hoad, C. L.; Marciani, L.; Ban, L.; Prayle, A. P.; Barr, H. L.; Jaudszus, A.; Mainz, J. G.; Spiller, R. C.; Gowland, P.; Major, G.; Smyth, A. R. Postprandial Changes in Gastrointestinal Function and Transit in Cystic Fibrosis Assessed by Magnetic Resonance Imaging. *J. Cyst. Fibros.* **2021**, *20* (4), 591–597. https://doi.org/10.1016/j.jcf.2020.06.004.

(9) Zhang, Y.; Jeon, M.; Rich, L. J.; Hong, H.; Geng, J.; Zhang, Y.; Shi, S.; Barnhart, T. E.; Alexandridis, P.; Huizinga, J. D.; Seshadri, M.; Cai, W.; Kim, C.; Lovell, J. F. Non-Invasive Multimodal Functional Imaging of the Intestine with Frozen Micellar Naphthalocyanines. *Nat. Nanotechnol.* **2014**, *9* (8), 631–638. https://doi.org/10.1038/nnano.2014.130.

(10) Zhong, Y.; Ma, Z.; Zhu, S.; Yue, J.; Zhang, M.; Antaris, A. L.; Yuan, J.; Cui, R.; Wan, H.; Zhou, Y.; Wang, W.; Huang, N. F.; Luo, J.; Hu, Z.; Dai, H. Boosting the Down-Shifting Luminescence of Rare-Earth Nanocrystals for Biological Imaging beyond 1500 nm. *Nat. Commun.* **2017**, *8* (1). https://doi.org/10.1038/s41467-017-00917-6.

(11) Yu, S.; Tu, D.; Lian, W.; Xu, J.; Chen, X. Lanthanide-Doped near-Infrared II Luminescent Nanoprobes for Bioapplications. *Sci. China Mater.* **2019**, *62* (8), 1071–1086. https://doi.org/10.1007/s40843-019-9414-4.

(12) Ling, S.; Yang, X.; Li, C.; Zhang, Y.; Yang, H.; Chen, G.; Wang, Q. Tumor Microenvironment-Activated NIR-II Nanotheranostic System for Precise Diagnosis and Treatment of Peritoneal Metastasis. *Angew. Chemie Int. Ed.* **2020**, *59* (18), 7219–7223. https://doi.org/10.1002/anie.202000947.

(13) Zhang, Y.; Yang, H.; An, X.; Wang, Z.; Yang, X.; Yu, M.; Zhang, R.; Sun, Z.; Wang, Q. Controlled Synthesis of $Ag_2Te@Ag_2S$ Core–Shell Quantum Dots with Enhanced and Tunable Fluorescence in the Second Near-Infrared Window. *Small* **2020**, *16* (14), 1–8. https://doi.org/10.1002/smll.202001003.

(14) Lian, W.; Tu, D.; Hu, P.; Song, X.; Gong, Z.; Chen, T.; Song, J.; Chen, Z.; Chen, X. Broadband Excitable NIR-II Luminescent Nano-Bioprobes Based on $CuInSe_2$ Quantum Dots for the Detection of Circulating Tumor Cells. *Nano Today* **2020**, *35*, 100943. https://doi.org/10.1016/j.nantod.2020.100943.

(15) Wang, T.; Wang, S.; Liu, Z.; He, Z.; Yu, P.; Zhao, M.; Zhang, H.; Lu, L.; Wang, Z.; Wang, Z.; Zhang, W.; Fan, Y.; Sun, C.; Zhao, D.; Liu, W.; Bünzli, J. C. G.; Zhang, F. A Hybrid Erbium(III)–Bacteriochlorin near-Infrared Probe for Multiplexed Biomedical Imaging. *Nat. Mater.* **2021**, *20* (11), 1571–1578. https://doi.org/10.1038/s41563-021-01063-7.

(16) Pei, P.; Chen, Y.; Sun, C.; Fan, Y.; Yang, Y.; Liu, X.; Lu, L.; Zhao, M.; Zhang, H.; Zhao, D.; Liu, X.; Zhang, F. X-Ray-Activated Persistent Luminescence Nanomaterials for NIR-II Imaging. *Nat. Nanotechnol.* **2021**, *16* (9), 1011–1018. https://doi.org/10.1038/s41565-021-00922-3.

(17) Chang, B.; Li, D.; Ren, Y.; Qu, C.; Shi, X.; Liu, R.; Liu, H.; Tian, J.; Hu, Z.; Sun, T.; Cheng, Z. A Phosphorescent Probe for in Vivo Imaging in the Second Near-Infrared Window. *Nat. Biomed. Eng.* https://doi.org/10.1038/s41551-021-00773-2.

(18) Xu, C.; Jiang, Y.; Han, Y.; Pu, K.; Zhang, R. A Polymer Multicellular Nanoengager for Synergistic NIR-II Photothermal Immunotherapy. *Adv. Mater.* **2021**, *2008061* (33), 1–11. https://doi.org/10.1002/adma.202008061.

(19) Muñoz-ortiz, T. Instantaneous in Vivo Imaging of Acute Myocardial Infarct by NIR-II Luminescent Nanodots. *Small* **2020**, *1907171*, 1–10. https://doi.org/10.1002/smll.201907171.

(20) Xu, W.; Wang, D.; Tang, B. Z. NIR-II AIEgens: A Win-Win Integration towards Bioapplications. *Angew. Chemie Int. Ed.* **2020**, *13(14)*, 7552-7563. https:// doi.org/10.1002/anie.202005899.

(21) Camacho-morales, R.; Rocco, D.; Xu, L.; Gili, F.; Dimitrov, N. Infrared Upconversion Imaging in Nonlinear Metasurfaces. Advanced Photonics **2021**, *3* (3), 1–10. https://doi.org/10.1117/1.AP.3.3.036002.

(22) Wang, R.; Zhou, L.; Wang, W.; Li, X.; Zhang, F. In Vivo Gastrointestinal Drug-Release Monitoring through Second near-Infrared Window Fluorescent Bioimaging with Orally Delivered Microcarriers. *Nat. Commun.* **2017**, *8*, 14702. https://doi.org/10.1038/ncomms14702.

(23) Lin, J.; Zeng, X.; Xiao, Y.; Tang, L.; Nong, J.; Liu, Y.; Zhou, H.; Ding, B.; Xu, F.; Tong, H.; Deng, Z.; Hong, X. Novel Near-Infrared II Aggregation-Induced Emission Dots for in Vivo Bioimaging. *Chem. Sci.* **2019**, *10* (4), 1219–1226. https://doi.org/10.1039/c8sc04363a.

(24) Wang, W.; Kong, Y.; Jiang, J.; Xie, Q.; Huang, Y.; Li, G.; Wu, D.; Zheng, H.; Gao, M.; Xu, S.; Pan, Y.; Li, W.; Ma, R.; Wu, M. X.; Li, X.; Zuilhof, H.; Cai, X.; Li, R. Engineering the Protein Corona Structure on Gold Nanoclusters Enables Red-Shifted Emissions in the Second Near-Infrared Window for Gastrointestinal Imaging. *Angew. Chemie Int. Ed.* **2020**, *59* (50), 22431–22435. https://doi.org/10.1002/anie.202010089.

(25) Wilson, R. H.; Nadeau, K. P.; Jaworski, F. B.; Tromberg, B. J.; Durkin, A. J. Review of Short-Wave Infrared Spectroscopy and Imaging Methods for Biological Tissue Characterization. *J. Biomed. Opt.* **2015**, *20* (3), 030901. https://doi.org/10.1117/1.jbo.20.3.030901.

(26) Li, B.; Zhao, M.; Feng, L.; Dou, C.; Ding, S.; Zhou, G.; Lu, L.; Zhang, H.; Chen, F.; Li, X.; Li, G.; Zhao, S.; Jiang, C.; Wang, Y.; Zhao, D.; Cheng, Y.; Zhang, F. Organic NIR-II Molecule with Long Blood Half-Life for in Vivo Dynamic Vascular Imaging. *Nat. Commun.* **2020**, *11*, 3102. https://doi.org/10.1038/s41467-020-16924-z.

(27) Fan, Y.; Wang, P.; Lu, Y.; Wang, R.; Zhou, L.; Zheng, X.; Li, X.; Piper, J. A.; Zhang, F. Lifetime-Engineered NIR-II Nanoparticles Unlock Multiplexed in Vivo Imaging. *Nat. Nanotechnol.* **2018**, *13* (10), 941–946. https://doi.org/10.1038/s41565-018-0221-0.

(28) Zhong, Y.; Ma, Z.; Wang, F.; Wang, X.; Yang, Y.; Liu, Y.; Zhao, X.; Li, J.; Du, H.; Zhang, M.; Cui, Q.; Zhu, S.; Sun, Q.; Wan, H.; Tian, Y.; Liu, Q.; Wang, W.; Garcia, K. C.; Dai, H. In Vivo Molecular Imaging for Immunotherapy Using Ultra-Bright near-Infrared-IIb Rare-Earth Nanoparticles. *Nat. Biotechnol.* **2019**, *37* (11),1322-1331. https://doi.org/10.1038/s41587-019-0262-4.

(29) Lei, X.; Li, R.; Tu, D.; Shang, X.; Liu, Y.; You, W.; Sun, C.; Zhang, F.; Chen, X. Intense Near-Infrared-II Luminescence from NaCeF$_4$:Er/Yb Nanoprobes for: In Vitro Bioassay and in Vivo Bioimaging. *Chem. Sci.* **2018**, *9* (20), 4682–4688. https://doi.org/10.1039/c8sc00927a.

(30) Zhou, J.; Chizhik, A. I.; Chu, S.; Jin, D. Single-Particle Spectroscopy for Functional Nanomaterials. *Nature* **2020**, *579* (7797), 41–50. https://doi.org/10.1038/s41586-020-2048-8.

(31) Liu, D.; Xu, X.; Du, Y.; Qin, X.; Zhang, Y.; Ma, C.; Wen, S.; Ren, W.; Goldys, E. M.; Piper, J. A.; Dou, S.; Liu, X.; Jin, D. Three-Dimensional Controlled Growth of Monodisperse Sub-50 nm Heterogeneous Nanocrystals. *Nat. Commun.* **2016**, *7*, 10254. https://doi.org/10.1038/ncomms10254.

(32) Yang, Y.; Mi, C.; Su, X.; Jiao, F.; Liu, L.; Zhang, J.; Yu, F.; Li, X.; Liu, Y.; Mai, Y. Ultraviolet C Upconversion Fluorescence of Trivalent Erbium in BaGd$_2$ZnO$_5$ Phosphor Excited by a Visible Commercial Light-Emitting Diode. *Opt. Lett.* **2014**, *39* (7), 2000. https://doi.org/10.1364/ol.39.002000.

(33) Chen, X.; Peng, D.; Ju, Q.; Wang, F. Photon Upconversion in Core-Shell Nanoparticles. *Chem. Soc. Rev.* **2015**, *44* (6), 1318–1330. https://doi.org/10.1039/c4cs00151f.

(34) Li, X.; Shen, D.; Yang, J.; Yao, C.; Che, R.; Zhang, F.; Zhao, D. Successive Layer-by-Layer Strategy for Multi-Shell Epitaxial Growth: Shell Thickness and Doping Position Dependence in Upconverting Optical Properties. *Chem. Mater.* **2013**, *25* (1), 106–112. https://doi.org/10.1021/cm3033498.

(35) Gu, Y.; Guo, Z.; Yuan, W.; Kong, M.; Liu, Y.; Liu, Y.; Gao, Y.; Feng, W.; Wang, F.; Zhou, J.; Jin, D.; Li, F. High-Sensitivity Imaging of Time-Domain near-Infrared Light Transducer. *Nat. Photonics* **2019**, *13* (8), 525–531. https://doi.org/10.1038/s41566-019-0437-z.

(36) Dong, N. N.; Pedroni, M.; Piccinelli, F.; Conti, G.; Sbarbati, A.; Ramírez-Hernández, J. E.; Maestro, L. M.; Iglesias-De La Cruz, M. C.; Sanz-Rodriguez, F.; Juarranz, A.; Chen, F.; Vetrone, F.; Capobianco, J. A.; Solé, J. G.; Bettinelli, M.; Jaque, D.; Speghini, A. NIR-to-NIR Two-Photon Excited $CaF_2:Tm^{3+},Yb^{3+}$ Nanoparticles: Multifunctional Nanoprobes for Highly Penetrating Fluorescence Bio-Imaging. *ACS Nano* **2011**, *5* (11), 8665–8671. https://doi.org/10.1021/nn202490m.

(37) Elinwa, A. U.; Umar, M. X-Ray Diffraction and Microstructure Studies of Gum Arabic-Cement Concrete. *Constr. Build. Mater.* **2017**, *156*, 632–638. https://doi.org/10.1016/j.conbuildmat.2017.08.162.

(38) Wang, P.; Fan, Y.; Lu, L.; Liu, L.; Fan, L.; Zhao, M.; Xie, Y.; Xu, C.; Zhang, F. NIR-II Nanoprobes in-Vivo Assembly to Improve Image-Guided Surgery for Metastatic Ovarian Cancer. *Nat. Commun.* **2018**, *9* (1), 2898. https://doi.org/10.1038/s41467-018-05113-8.

(39) Tian, R.; Ma, H.; Zhu, S.; Lau, J.; Ma, R.; Liu, Y.; Lin, L.; Chandra, S.; Wang, S.; Zhu, X.; Deng, H.; Niu, G.; Zhang, M.; Antaris, A. L.; Hettie, K. S.; Yang, B.; Liang, Y.; Chen, X. Multiplexed NIR-II Probes for Lymph Node-Invaded Cancer Detection and Imaging-Guided Surgery. *Adv. Mater.* **2020**, *32* (11), 1–10. https://doi.org/10.1002/adma.201907365.

(40) Wang, F.; Wan, H.; Ma, Z.; Zhong, Y.; Sun, Q.; Tian, Y.; Qu, L.; Du, H.; Zhang, M.; Li, L.; Ma, H.; Luo, J.; Liang, Y.; Li, W. J.; Hong, G.; Liu, L.; Dai, H. Light-Sheet Microscopy in the near-Infrared II Window. *Nat. Methods* **2019**, *16* (6), 545–552. https://doi.org/10.1038/s41592-019-0398-7.

(41) Liu, S.; Zhu, Y.; Wu, P.; Xiong, H. Highly Sensitive D-A-D-Type Near-Infrared Fluorescent Probe for Nitric Oxide Real-Time Imaging in Inflammatory Bowel Disease. *Anal. Chem.* **2021**, *93* (11), 4975–4983. https://doi.org/10.1021/acs.analchem.1c00281.

(42) Finnberg, N. K.; Liu, Y.; El-Deiry, W. S. Detection of DSS-Induced Gastrointestinal Mucositis in Mice by Non-Invasive Optical near-Infrared (NIR) Imaging of Cathepsin-Activity. *Cancer Biol. Ther.* **2013**, *14* (8), 736–741. https://doi.org/10.4161/cbt.25094.

(43) Lee, A.; De Mei, C.; Fereira, M.; Marotta, R.; Yoon, H. Y.; Kim, K.; Kwon, I. C.; Decuzzi, P. Dexamethasone-Loaded Polymeric Nanoconstructs for Monitoring and Treating Inflammatory Bowel Disease. *Theranostics* **2017**, *7* (15), 3653–3666. https://doi.org/10.7150/thno.18183.